\documentclass[prd,twocolumn,showpacs,amsmath,amssymb,superscriptaddress,preprintnumbers,nofootinbib]{revtex4}
\usepackage{amssymb}
\usepackage{amsmath}
\usepackage{graphicx}

\begin{document}
\title{New application of the Killing vector field formalism: \\ Modified periodic potential and two-level profiles \\ of the axionic dark matter distribution}

\author{Alexander B. Balakin}
\email{Alexander.Balakin@kpfu.ru} \affiliation{Department of
General Relativity and Gravitation, Institute of Physics, Kazan
Federal University, Kremlevskaya str. 16a, Kazan 420008, Russia}

\author{Dmitry E. Groshev}
\email{groshevdmitri@mail.ru} \affiliation{Department of
General Relativity and Gravitation, Institute of Physics, Kazan
Federal University, Kremlevskaya str. 16a, Kazan 420008, Russia}


\begin{abstract}
We consider the structure of halos of the axionic dark matter, which surround  massive relativistic objects  with static spherically symmetric gravitational field and monopole-type magneto-electric fields.  We work with the model of pseudoscalar field with the extended periodic potential, which depends on additional arguments proportional to the moduli of the Killing vectors; in our approach they play the roles of model guiding functions. The covariant model of the axion field with this modified potential is equipped  with the extended formalism of the Killing vector fields, which is established in analogy with the formalism of the Einstein-aether theory, based on the introduction of a unit timelike dynamic vector field. We study the equilibrium state of the axion field, for which the extended potential and its derivative vanish, and illustrate the established formalism by the analysis of two-level axionic dark matter profiles, for which the stage delimiters relate to the critical values of the modulus of the timelike Killing vector field.
\end{abstract}


\maketitle

\section{Introduction}

Dark matter is one of the key elements of modern scenaria of the Universe evolution, of the galactic formation and rotation \cite{DM1,DM2,DM3}.
Dark matter forms halos and subhalos, filaments and walls; their structures give a significant part of information about the gravitational field of the central bodies, which are surrounded by these
galactic units. It is impossible to observe dark matter directly, since it neither emits, nor scatters light, that is why theoretical predictions based on the structure modeling of the dark matter halos, subhalos, filaments and walls play the important role in the investigations of these enigmatic cosmological units.

We follow the line that the axions, massive pseudo-Goldstone bosons, predicted in the works \cite{A1,A2,A3}, are the particles, which are responsible for the phenomena, associated with the dark matter. During the last forty years the basic aspects of the axion theory, the cosmological and astrophysical applications of the axion models were well documented in many works (see, e.g., \cite{Rev1,Rev3,Rev4,Rev5,Rev6,Rev7,Rev8,Rev10,Rev11,Rev12}). During the last decade new trends in the theory of the axionic dark matter appeared. First of all, we mean the non-minimal extensions of the axion models, which are based on the $F(R)$ modifications of the theory of gravity (see, e.g., \cite{NM1,NM2,NM3,NM4,NM5}), on the non-minimal extensions of the theory of the axion-photon coupling \cite{NM6}, and on the axionically extended models of the dynamic aether \cite{NM7}.
The second trend is associated with the recent surge of interest to the black hole mergers and supermassive black holes: the axionic dark matter can play the role of a marker revealing specific features of the strong gravitational field (see, e.g., \cite{BH1,BH2} as illustrations).
The third trend is connected with the modeling of axionic structures of different scales, associated with the galactic halos, miniclusters, halos surrounding the magnetars, dyons, etc. (see, e.g., \cite{H1,H3,BG1,BG2}). We have to emphasize, that one of the most influential idea in these new models is the idea of internal self-interaction in the axionic systems. The axionic dark matter, apparently, is not a simple collisionless gas without pressure, which interacts by the gravitational field only; probably, the axions form correlated systems. The most known model of this type is the model of axionic Bose-Einstein condensate, presented in \cite{BEC}. The alternative description of correlated axionic systems is based on the models containing the potential of the pseudoscalar field $V(\phi^2)$, which can have the polynomial form $\phi^n$, the $\phi^4$ Higgs - type structure, or the nonlinear  periodic form (see, e.g., \cite{H2}).

We focus on the description of the axionic systems based on the analysis of the equation for the pseudoscalar field $\phi$, associated with axions \cite{A8,A9}.
New aspects of this theory are connected with the structure of the pseudoscalar field potential $V$. The standard potential is considered as an even function of the argument $\phi$ only, i.e., $V=V(\phi^2)$; we extend the theory by introducing the function of $s+1$ arguments $V(\phi^2, \xi_{(1)},\xi_{(2)}, ...\xi_{(s)})$, where $\left\{\xi_{(a)} \right\}$ is the set of some scalars ($a{=}1,2,...s$). What can be the origin of these scalars?

\noindent
{\it (i) Scalars associated with the Einstein-aether theory.}

\noindent
The extension of the potential was prompted by the analogy with the Einstein-aether theory \cite{EA1,EA2,EA3}, which is based on the introduction of a timelike dynamic vector field $U^i$ normalized by unity ($g_{ik}U^iU^k=1$). In the framework of the Einstein-aether theory one can introduce into the potential $V$ four auxiliary scalars based on the decomposition of the covariant derivative $\nabla_mU_n$ (see, e.g., \cite{S1,S2,S3,S4} for details and references). The first scalar is the divergence of the vector field ($\xi_{(1)} \to \Theta \equiv \nabla_m U^m$). The second scalar is the square of the acceleration four-vector $a^i = U^k \nabla_k U^i$ ($\xi_{(2)} \to a^2 \equiv g_{ik} a^ia^k$). The third scalar is the square of the symmetric shear tensor $\sigma_{pq}$ ($\xi_{(3)} \to \sigma^2 \equiv \sigma_{ik} \sigma^{ik}$), and the fourth scalar is constructed using the square of the skew-symmetric vorticity tensor $\omega_{ik}$ ($\xi_{(4)} \to \omega^2 \equiv \omega_{ik}\omega^{ik}$). The scalar coinciding with the modulus of the vector $U^i$ is equal to one and thus it can not be considered as a guiding function.   In other words, when the theory possesses intrinsic unit vector field, we can add four new arguments to the potential of the pseudoscalar field $V$, and can use this covariant extension for detailing the structure of the dark matter configuration.

There are three illustrations of this idea. The first one relates to the Friedmann type cosmology, for which $a^i =0$, $\sigma_{ik}=0$, $\omega_{ik}=0$, and the only scalar $\Theta$ proportional to the Hubble function $H(t)$ ($\Theta = 3H$) is the nonvanishing guiding function (see, e.g., \cite{B1,B2}). The second illustration is connected with the pp-wave symmetric spacetimes with two auxiliary nonvanishing scalars: $\Theta$ and $\sigma^2$ \cite{B3,B4}. The third example appeared in the G\"odel type model, in the framework of which  the nonvanishing parameter $\omega^2$ can be considered as the guiding function \cite{B5}.

\noindent
{\it (ii) Scalars associated with the Killing vectors.}

\noindent
If the model does not contain vector field $U^i$ associated with the dynamic aether, but possesses some specific symmetries, described by the set of Killing vectors and/or conformal Killing vectors $\xi^i_{(a)}$ ($a=1,2,...10$), we suggest to use the new scalar quantities of two types. First, we can consider the moduli of the Killing vectors, $\xi_{(a)}= \sqrt{|g_{mn} \xi^m_{(a)}\xi^n_{(a)}|}$ (or their combinations) as the auxiliary arguments of the modified axionic potential. It was impossible in the Einstein-aether theory, since the modulus of the velocity field $U^i$ is equal to one. The scalar quantities of the second type contain nonvanishing convolutions of the derivative $\nabla_m \xi_{n}$. When we deal with the standard Killing vector, which satisfies the equation $\nabla_m \xi_{n}{+}\nabla_n \xi_{m}=0$, the divergence $\nabla_m \xi^{m}$ is equal to zero, but for the so-called conformal Killing vector, which satisfies the equation $\nabla_m \xi_{n}{+}\nabla_n \xi_{m}= \frac12 g_{mn} \nabla_s \xi^{s}$, the scalar $\nabla_s \xi^{s}$ is nonvanishing and can be used for the physical modeling. In its turn, for the standard Killing vector the skew-symmetric quantity $\frac12 \left[\nabla_m \xi_{n}{-}\nabla_n \xi_{m}\right]= \nabla_m \xi_{n}$ is nonvanishing, and its square can be used in analogy with the square of vorticity tensor $\omega_{mn}$ appeared in the Einstein-aether theory. Illustration of this idea can be found in \cite{BG1}, where the model with the static spherically symmetric spacetime of the Reissner-Nordstr\"om type is analyzed.

\noindent
{\it (iii) Why the new guiding scalar functions are necessary for the extended analysis?}

\noindent

In order to describe the new units in the dark matter halos (filaments, walls, etc.), we need of covariant scheme to fix some lines and surfaces.
For instance, when we deal with a static spherically symmetric gravitational field and try to separate one spatial domain from the other, we could use the well-known Heaviside function $h$; however, we can not insert into the Lagrangian the term $h(r{-}r_*)$, since the difference of radial coordinates is non-invariant quantity. Nevertheless, we can use the term $h(\xi-\xi_*)$, where $\xi$ is the appropriate additional scalar, and $\xi_*$ is its critical value; this scheme is the covariant one.

\noindent
{\it (iv) How the paper is organized?}

\noindent
In order to provide the self-consistency of the approach, which is based on the extended formalism of Killing vectors, we have to introduce the Killing vector field as the dynamic field, i.e., we have to add into the Lagrangian the scalar terms quadratic in the covariant derivative $\nabla_m \xi_n$, and to develop the variation formalism in analogy with the formalism used in the Einstein-aether theory. In Section IIA of this work we establish the strict covariant formalism justifying this idea in general case.

Since the additional scalars, based on the Killing vector fields, are designed to be used for description of the pseudoscalar (axion) field, we have to modify the axionic  periodic potential and to derive the correspondingly modified master equations for the axion, electromagnetic and gravitational fields; in Section IIB we present these modified master equations and discuss the concept of the equilibrium state of the axionic subsystem.

In Section III we apply the formalism for description of two-level distribution of the axionic dark matter near the static spherically symmetric objects with magnetic and electric fields; the analysis of the obtained key equation is presented in Section IV. In Section V we discuss the features of the profiles of the axionic dark matter obtained in the framework of the suggested model.

\section{Extended formalism, which includes Killing vector fields}

\subsection{Analogy with the Einstein-aether theory}

\subsubsection{Action functional}

The  action functional of the Einstein-aether theory is known to have the form
$$
S_{(\rm EA)} = \int d^4 x \sqrt{-g} \ \frac{1}{2\kappa}\left\{R+2\Lambda) + \lambda(U^mU_m-1) + \right.
$$
\begin{equation}
\left. + K^{abmn}\nabla_a U_{m} \nabla_b U_n \right\} \,,
\label{1}
\end{equation}
where $R$ is the Ricci scalar; $\Lambda$ is the cosmological constant; $U^i$ is the vector field, $\nabla_k$ is the covariant derivative;  $K^{abmn}$  is the Jacobson's constitutive tensor:
\begin{equation}
K^{abmn}{=} C_1 g^{ab}g^{mn} {+} C_2 g^{am}g^{bn} {+} C_3 g^{an}g^{bm} {+} C_4 U^a U^b g^{mn},
\label{2}
\end{equation}
which contains four phenomenological constants $C_1$, $C_2$, $C_3$ and $C_4$. The scalar quantity $\lambda$ is the Lagrange multiplier; variation with respect to $\lambda$ gives the constraint
$g_{mn}U^mU^n=1$; this quadratic relationship is, in fact, the normalization condition for the vector field.

When the vector field $U^i$ is absent, but there exists the Killing vector field $\xi^i$, we suggest to work with the action functional similar to (\ref{1}):
$$
S_{(\rm EK)} = \int d^4 x \sqrt{-g} \ \frac{1}{2\kappa}\left\{R+2\Lambda) + \right.
$$
\begin{equation}
\left. {+} \tilde{\lambda} \left[\nabla_{(k}\xi_{m)} \nabla^{(k}\xi^{m)}{-} \frac14 (\nabla_n \xi^n)^2 \right]{+} {\cal K}^{abmn}\nabla_a \xi_{m} \nabla_b \xi_n \right\},
\label{3}
\end{equation}
where the parentheses  denote symmetrization: $\nabla_{(i}\xi_{m)} {=} \frac12 \left(\nabla_{i}\xi_{m} {+} \nabla_{m}\xi_{i} \right)$, and the tensor ${\cal K}^{abmn}$
\begin{equation}
{\cal K}^{abmn} = {\cal K}_1 g^{ab}g^{mn} + {\cal K}_2 g^{am}g^{bn} + {\cal K}_3 g^{an}g^{bm}
\label{4}
\end{equation}
contains only three phenomenological constants ${\cal K}_1$, ${\cal K}_2$ and ${\cal K}_3$ (we assume that the constitutive tensor ${\cal K}^{abmn}$ contains the metric tensor, but does not include the Killing vector itself). In other words, we consider the vector field $\xi^i$ as a dynamic quantity in analogy with the aether velocity field, however, instead of the algebraic constraint we use the differential one.

\subsubsection{Variation with respect to $\tilde{\lambda}$ and Killing equations}

Variation of the action functional (\ref{3}) with respect to the Lagrange multiplier $\tilde{\lambda}$ gives the constraint
\begin{equation}
\nabla_{(k}\xi_{m)} \nabla^{(k}\xi^{m)} - \frac14 (\nabla_n \xi^n)^2 = 0  \,,
\label{5}
\end{equation}
which can be rewritten as follows:
\begin{equation}
\left[\nabla_{(k}\xi_{m)} {-} \frac14 g_{km} \nabla_n \xi^n \right] \left[\nabla^{(k}\xi^{m)} {-} \frac14 g^{km} \nabla_s \xi^s \right]= 0  \,.
\label{6}
\end{equation}
This equation is satisfied, if
\begin{equation}
\nabla_{k}\xi_{m} {+} \nabla_{m}\xi_{k} = \frac12 g_{km} \nabla_n \xi^n \,.
\label{7}
\end{equation}
Generally speaking, (\ref{7}) can be classified as the sufficient but not the necessary condition for the fulfillment of the quadratic relationship (\ref{6}), nevertheless, here we do not discuss this fine mathematical detail.

Clearly, (\ref{7}) are the equations for the so-called conformal Killing vector, if $\nabla_n \xi^n \neq 0$, and for the standard Killing vector, if $\nabla_n \xi^n =0$ (see, e.g., \cite{Exact} for details). Calculation of the divergence of the left and right-hand sides of (\ref{7}) yields
\begin{equation}
\nabla_m \nabla^m \xi_k + R_{kj} \xi^j = -  \frac12 \nabla_k (\nabla_n \xi^n) \,.
\label{8}
\end{equation}
Also, when we deal with the standard Killing vector field (i.e., $\nabla_k \xi^k=0$) we can use the relationships
\begin{equation}
\nabla_s \nabla_m \xi_k + R_{jskm} \xi^j =0 \,,
\label{8999}
\end{equation}
as the integrability condition of the first order of the Killing equation.
Here, as usual, $R_{kj}= R^m_{\ \ kmj}$ is the Ricci tensor, and $R^i_{\ kmj}$  is the Riemann tensor.

\subsubsection{Variation with respect to $\xi^i$}

When one deals with the Einstein-aether theory, the variation with respect to the velocity four-vector $U^i$ gives the master equation
\begin{equation}
\nabla_a \left[K^{abjn} \nabla_b U_n \right] = \lambda U^j + C_4 U^b \nabla_bU_n \nabla^j U^n \,.
\label{9}
\end{equation}
Thus, the normalization condition yields
\begin{equation}
\lambda = U_m \nabla_a \left[K^{abmn} \nabla_b U_n \right] - C_4 U^b \nabla_bU_n  \ U^a \nabla_a U^n \,.
\label{10}
\end{equation}
When we deal with the extended formalism, which includes the Killing vector field, variation of the functional (\ref{3}) with respect to $\xi^j$ gives the equations
\begin{equation}
\nabla_a \left\{\tilde{\lambda} \left[\nabla^{(a} \xi^{m)}- \frac14 \nabla_n \xi^n \right] + 2{\cal K}^{abjn} \nabla_b \xi_n \right\}=0 \,.
\label{11}
\end{equation}
Keeping in mind the relationships (\ref{4}), (\ref{7}) and (\ref{8}) we can reduce these equations to the relations
\begin{equation}
\left({\cal K}_3 {-} {\cal K}_1 \right) R_{js} \xi^s {+} \left({\cal K}_2 {+} {\cal K}_3 {-} \frac12{\cal K}_1  \right)\nabla_j \nabla_m \xi^m  =0 \,,
\label{12}
\end{equation}
which does not include $\tilde{\lambda}$. The equation (\ref{12}) is satisfied identically, when
\begin{equation}
{\cal K}_2 = - \frac12 {\cal K}_1 \,, \quad {\cal K}_3 = {\cal K}_1
\label{119}
\end{equation}
for arbitrary conformal Killing vector with $\nabla_n \xi^n \neq 0$. When we deal with the standard Killing vector, i.e., $\nabla_n \xi^n {=} 0$, the parameter ${\cal K}_2$
remains arbitrary and hidden.

\subsubsection{Variation with respect to metric}

When one deals with the Einstein-aether theory the contribution of the vector field into the total stress-energy tensor is known to have the following form
$$
T^{({\rm U})}_{ik} =
\frac12 g_{ik} {\cal J}^{am} \nabla_a U_m {+}
$$
$$
{+}\nabla^m \left[U_{(i}{\cal J}_{k)m}\right] {-}
\nabla^m \left[{\cal J}_{m(i} U_{k)} \right] {-}
\nabla_m \left[{\cal J}_{(ik)} U^m\right]+
$$
$$
+C_1\left[(\nabla_mU_i)(\nabla^m U_k) {-}
(\nabla_i U_m \nabla_k U^m) \right] {+}
$$
\begin{equation}
{+}C_4 (U^a \nabla_a U_i)(U^b \nabla_b U_k) \,,
\label{13}
\end{equation}
where the following definition is used:
\begin{equation}
{\cal J}^{am} = K^{abmn} \nabla_b U_n \,.
\label{14}
\end{equation}
In order to obtain the similar formula for our extended model, let us find the corresponding analogies.
We assume now that ${\cal K}_1 = {\cal K}_3$, ${\cal K}_2 = - \frac12 {\cal K}_1$, and obtain that
\begin{equation}
{\cal K}^{abmn} = {\cal K}_1 \left(g^{ab}g^{mn} + g^{an}g^{bm} - \frac12 g^{am}g^{bn} \right) \,,
\label{15}
\end{equation}
\begin{equation}
{\cal K}^{abmn} \nabla_b \xi_n = 2 {\cal K}_1 \left[ \nabla^{(a} \xi^{m)} - \frac14g^{am} \nabla_n \xi^n \right]\,,
\label{16}
\end{equation}
\begin{equation}
{\cal K}^{abmn} \nabla_a \xi_m \nabla_b \xi_n = 2 {\cal K}_1 \left[\nabla^{(a} \xi^{m)} \nabla_{(a} \xi_{m)}{-} \frac14 \left(\nabla_n \xi^n \right)^2 \right].
\label{17}
\end{equation}
This means that the corresponding term in the integral (\ref{3}) can be rewritten as follows
\begin{equation}
\lambda^{*} \left[\nabla^{(a} \xi^{m)} \nabla_{(a} \xi_{m)} - \frac14 \nabla_n \xi^n \right]\,, \quad \lambda^{*} \equiv (\tilde{\lambda}+ 2{\cal K}_1) \,.
\label{18}
\end{equation}
Thus, the analogies are the following: first, we have to omit $C_4$; second, the analog of the term  ${\cal J}^{am}$ is symmetric and it vanishes on the solutions to the Killing equations. In other words, following this analogy we obtain the vanishing stress-energy tensor associated with the Killing vector field.

If we fulfil the direct variation of the action functional (\ref{3}) with respect to metric, the corresponding effective stress-energy tensor $T_{pq}^{(\rm K)}$ associated with the contribution of the Killing vector field can be written as follows:
$$
T_{pq}^{(\rm K)} = \frac{(-2)}{\sqrt{-g}}\frac{\delta}{\delta g^{pq}} \left\{\sqrt{{-}g}\lambda^{*}  g^{ab} g^{mn} \times   \right.
$$
\begin{equation}
 \left.  \left[\nabla_{(a} \xi_{m)} {-} \frac14 g_{am}(\nabla_s \xi^s) \right] \left[\nabla_{(b} \xi_{n)} {-} \frac14 g_{bn}(\nabla_l \xi^l) \right]\right\}.
\label{19}
\end{equation}
Using (\ref{7}) and (\ref{8}), we can show that this tensor takes zero value, $T_{pq}^{(\rm K)} = 0$. This means that in the suggested scheme there are no additional contributions  to the equations of the gravitational field associated with conformal and/ot standard  Killing vector fields.

\subsubsection{Short summary}

1. If we consider the proposed scheme of using of the Killing vector field as a dynamic field based on the action functional (\ref{3}) with coupling parameters (\ref{119}), constraint (\ref{7}) and its differential consequence (\ref{8}), we obtain non-violated equations for the gravitational field.

2. If we consider the s-parameter group and deal with the set of Killing vector fields $\left\{\xi_{(a)} \right\}$ ($a=1,2,..s$),  we can extend the model by inserting the term  $\sum_{a} \lambda^{*}_{(a)}\left[ \nabla_{(m} \xi_{(a)n)} \nabla^{(m} \xi_{(a)}^{n)} - \frac14 (\nabla_n \xi^n_{(a)})^2\right]$ instead of $\lambda^{*}\left[\nabla_{(m} \xi_{n)} \nabla^{(m} \xi^{n)}-\frac14 (\nabla_n \xi^n)^2 \right]$ into the Lagrangian.

3. For the spatially homogeneous cosmological models of the Friedmann type with the scale factor $a(t)$ there are one conformal time-like Killing vector $\xi^i_{(0)} = a(t) \delta^i_0$, and three space-like divergence-free  Killing vectors $\xi^i_{(\alpha)} = \delta^i_{\alpha}$, where $\alpha = 1,2,3$. Clearly, using the moduli of these Killing vectors we can construct only one additional scalar $\xi = a(t)$, or can choose more convenient quantity $x=\frac{a(t)}{a(t_0)}$ (see, e.g., \cite{BB1,BB2,BB3}).

4. When we consider a static model, we can use the time-like Killing vector $\xi^i_{(0)} {=} \delta^i_0$, so that its modulus $\xi_{(0)} {=} \sqrt{g_{00}}$ can play the role of the guiding function (see \cite{BG1}).

5. For spherically symmetric  models we can take the azimuthal Killing vector $\xi^i_{(\varphi)} = \delta^{i}_{\varphi} $ to obtain the additional scalar $\xi_{(\varphi)} = r \sin{\theta} $, where, as usual,
$r$ is the radial variable, $\theta$ is the meridional angle, $\varphi$ is the azimuthal angle.

6. When we deal with the spacetimes with the so-called pp-wave symmetry, we obtain one covariant constant null Killing vector $\xi^i_{(v)} {=} \delta_0^i {-} \delta_1^i$ and two space-like Killing vectors $\xi^i_{(2)} {=} \delta_2^i$ and $\xi^i_{(3)} {=} \delta_3^i$. Thus, we can use two moduli $\xi_{(2)} {=} \sqrt{|g_{22}|}$ and $\xi_{(3)} {=} \sqrt{|g_{33}|}$, as well as, the scalar product $g_{ik}\xi^i_{(2)}\xi^k_{(3)} {=} g_{23}$, as additional guiding functions.

\subsection{Total action functional and extended master equations for interacting fields}

\subsubsection{Extended action functional}

Let us consider now the total action functional
$$
S_{(\rm total)} =
\int d^4 x \sqrt{-g} \left\{ L^{({\rm matter})} + \frac{1}{2\kappa}\left(R+2\Lambda \right) + \right.
$$
$$
\left. + \frac{1}{2\kappa} \sum_{a} \lambda^{*}_{(a)}\left[ \nabla_{(m} \xi_{(a)n)} \nabla^{(m} \xi_{(a)}^{n)} - \frac14 (\nabla_n \xi^n_{(a)})^2 \right]  + \right.
$$
\begin{equation}
\left. + \frac14 \left(F_{mn} {+} \phi F^{*}_{mn}\right) F^{mn} - \frac12  \Psi_{0}^2\left(
\nabla_m \phi \nabla^m \phi {-} V \right) \right\} \,.
\label{20}
\end{equation}
It describes the electromagnetic field, represented in terms of the Maxwell tensor $F_{mn}$ and its dual $F^{*}_{mn}$, which interacts with the pseudoscalar (axion) field $\phi$ with the potential $V$; $L^{({\rm matter})}$ is the Lagrangian of a baryonic matter; the quantity $\Psi_0$ is reciprocal to the coupling constant of the axion-photon interaction $\frac{1}{\Psi_0}{=} g_{A \gamma \gamma}$ (see, e.g., \cite{CAST} for its observational constraints).

\subsubsection{Ansatz about an equilibrium state of the axion system}

We assume now that the potential of the pseudoscalar field has the periodic structure
\begin{equation}
V(\phi, \xi_{(1)}, ... \xi_{(s)}) = \frac{m^2_A \Phi^2_{*}}{2\pi^2} \left[1- \cos{\left(\frac{2 \pi \phi}{\Phi_*}\right)} \right] \,,
\label{21}
\end{equation}
where $\Phi_*=\Phi_*(\xi_{(1)}, ... \xi_{(s)})$ is a function of the moduli of the Killing vectors $\xi_{(a)}$.
When $\phi = n \Phi_*$ with arbitrary integer $n$, the potential and its derivative take zero values
\begin{equation}
V_{|\phi=n\Phi_*} = 0 \,, \quad \left(\frac{\partial V}{\partial \phi}\right)_{|\phi=n\Phi_*} =0 \,,
\label{22}
\end{equation}
the values $\phi=n\Phi_*$ correspond to minima of the potential. The coefficient in front of the periodic function in (\ref{21}) is chosen so that, when $\phi$ has a small deviation from the minimum value $\phi=n\Phi_*$ (i.e., when $\phi=n\Phi_* {+} \psi$ and $|\psi|<<1$),  the potential converts into $V \to  m^2_A \psi^2$.
Keeping in mind the mechanical analogy that equilibrium states of dynamic systems can be realized just in the minimum of the corresponding potential, we use below the special term {\it equilibrium state} of the axion field, $\phi_{(\rm eq)}$, if the potential of the pseudoscalar field $\phi$ and its derivative with respect to $\phi$ takes zero value at $\phi = \phi_{(\rm eq)}$.

\subsubsection{Extended master equations}

The variation procedure  gives the system of master equations of the model, which contains four sub-sets.

\vspace{2mm}
{\it (i)}
The first sub-set appears as the result of variation of the total functional (\ref{20}) with respect to the electromagnetic potential $A_i$; it includes the extended Maxwell equations
\begin{equation}
\nabla_k \left[F^{ik} {+} \phi  F^{*ik} \right] = 0 \,,
\label{23}
\end{equation}
\begin{equation}
 F_{ik} = \nabla_i A_k - \nabla_k A_i \ \Rightarrow  \nabla_k F^{ik*} =0 \,.
 \label{239}
\end{equation}
When the axion field is in the equilibrium state, we have to replace $\phi$ with $n \Phi_{*}$ in (\ref{23}).

\vspace{2mm}
{\it (ii)}
The second sub-set consists of one equation for the pseudoscalar field:
\begin{equation}
\nabla_k \nabla^k \phi + \frac12 \frac{\partial V}{\partial \phi} = -
\frac{1}{4\Psi_{0}^2} F^{*}_{mn}F^{mn} \,,
\label{24}
\end{equation}
it is the result of variation of the total action functional with respect to $\phi$. When the axion field is in the equilibrium state this equation converts into
\begin{equation}
\nabla_k \nabla^k \Phi_{*} = - \frac{1}{4 n \Psi_{0}^2} F^{*}_{mn}F^{mn} \,.
\label{249}
\end{equation}

\vspace{2mm}
{\it (iii)}
The third subset appears as the result of variation with respect to the vector fields $\xi^i_{(a)}$; it has the form
$$
\nabla_k \left\{\lambda^{*}_{(a)} \left[\nabla^{(k} \xi^{m)}_{(a)} - \frac14 g^{km} (\nabla_n \xi^n_{(a)})^2 \right]\right\} =
$$
\begin{equation}
= - \kappa \Psi^2_0 \left(\frac{\partial V}{\partial \Phi_{*}}\right) \left(\frac{\partial \Phi_{*}}{\partial \xi_{(a)}}\right) \left(\frac{\xi^m_{(a)}}{\xi_{(a)}}\right)   \,.
\label{25}
\end{equation}
When we deal with the standard and/or conformal Killing vectors $\xi_{(a)}$, the left-hand side of this equation vanishes, as it was shown in Section IIA. When the axion field is in the equilibrium state, the quantity $\frac{\partial V}{\partial \Phi_{*}}$ (see (\ref{21})) also is equal to zero.
In this sense, our scheme of extension of the potential of the pseudoscalar (axion) field using the moduli of the Killing vectors is self-consistent, when we assume that the axion field takes one of the equilibrium values
$\pm \Phi_*, \pm 2\Phi_*, ..., \pm n \Phi_*, ...$. In other words, the equations for the fields $\xi_{(a)}$ coincide with the Killing equations (\ref{7}).

\vspace{2mm}
{\it (iv)}
Variation with respect to metric gives the equations for the gravitational field in the following form:
\begin{equation}
\frac{1}{\kappa}\left[R_{ik} {-} \frac12 R g_{ik} {-} \Lambda g_{ik} \right] {=} T^{({\rm m})}_{ik} {+} T^{({\rm axion})}_{ik}{+} T^{({\rm em})}_{ik} {+} T^{({\rm V})}_{ik}.
\label{26}
\end{equation}
Here the stress-energy tensors  of the matter, of the pseudoscalar (axion) field and of the electromagnetic field are given, respectively,  by
\begin{equation}
T^{({\rm m})}_{ik} = -\frac{2}{\sqrt{-g}} \frac{\delta}{\delta g^{ik}}\left[\sqrt{-g} L^{{({\rm matter})}} \right]\,,
\label{27}
\end{equation}
\begin{equation}
T^{({\rm axion})}_{ik}= \Psi^2_0 \left[\nabla_i\phi\nabla_k\phi-\frac12 g_{ik} \left(\nabla_n \phi \nabla^n \phi-V \right) \right] \,, \label{28}
\end{equation}
\begin{equation}
T^{({\rm em})}_{ik}=\frac{1}{4}g_{ik}F_{mn}F^{mn}-F_{in}{F_{k}}^n \,.
\label{29}
\end{equation}
The last term relates to the contribution of the interaction between the axion field and Killing vector field; it can be written as follows:
\begin{equation}
T^{({\rm V})}_{ik} = \frac12 \Psi^2_0 \left(\frac{\partial V}{\partial \Phi_*}\right) \sum_{a}\frac{\xi_{i(a)}\xi_{k(a)}}{\xi_{(a)}} \left(\frac{\partial \Phi_*}{\partial \xi_{(a)}} \right) \,.
\label{30}
\end{equation}
The term $\frac{\partial V}{\partial \Phi_*}$ can be presented in the form:
\begin{equation}
\frac{\partial V}{\partial \Phi_*} = \frac{2}{\Phi_{*}} V - \frac{m^2_A \phi}{\pi} \  \sin{\left(\frac{2 \pi \phi}{\Phi_*}\right)}  \,.
\label{217}
\end{equation}
Clearly, when the axion field is in the equilibrium state, i.e., $\phi = n \Phi_*$, the term $T^{({\rm V})}_{ik}$ vanishes, and the gravitational field equations remain non-violated.

\section{Master equations for the model of static spherically symmetric gravitational field}

\subsection{Geometry of the outer zone and representation of the guiding functions}

Let us consider now an outer zone of a static spherically symmetric object, which possesses a magnetic charge $\mu$ and an electric charge $q$; the term $|Q|\equiv \sqrt{\mu^2{+} q^2}$ describes the hybrid charge. The mentioned object can be presented, for instance, by a charged black hole; in this case the outer zone covers the spacetime domain from the outer black hole horizon to the infinity. When we consider a dyon, we deal with the zone, which covers the domain from the boundary of the solid body of this object to the infinity. We assume that in both cases the metric in the outer zone can be  presented as follows:
\begin{equation}
ds^2 = N(r)dt^2 -\frac{dr^2}{N(r)} - r^2 \left(d\theta^2 + \sin^2{\theta} d\varphi^2 \right) \,,
\label{31}
\end{equation}
and the metric coefficient $N(r)$ is effectively described by the Reissner-Nordstr\"om function
\begin{equation}
N=1-\frac{2M}{r} + \frac{{Q^2}}{r^2} \,.
\label{32}
\end{equation}

\subsubsection{First guiding function, $\xi$}

The static spacetime with the metric (\ref{31}) is known to admit the existence of the time-like Killing vector field $\xi^i{=}\delta^i_0$, and the modulus of this four-vector is
$\xi \equiv \sqrt{N(r)}$. The value of the modulus $\xi$ on the outer Reissner-Nordstr\"om horizon $r_{+}= M {+} \sqrt{M^2{-}Q^2}$ is equal to zero, $\xi(r_+) {=}0$, and its value at infinity is equal to one, $\xi(\infty) {=}1$. In other words, the scalar $\xi$ takes values in the interval $(0,1)$ and can be used as the invariant delimiter of the first type. This means that if one needs to distinguish the specific sphere $r{=}r_{*}$, one can use this first scalar $\xi$ and the appropriate delimiter $\xi {=}\xi_{*}{=}\sqrt{1-\frac{2M}{r_*}{+}\frac{Q^2}{r^2_{*}}}$.

\subsubsection{Second guiding function, $\eta$}

The spacetime with the metric (\ref{31}) admits the so-called azimuthal Killing vector $\xi^i_{(\varphi)} = \delta^i_{\varphi}$, whose modulus is $\xi_{(\varphi)}= r \sin{\theta}$. Based on this fact we introduce the transformed scalar quantity
\begin{equation}
\eta \equiv \arcsin{\frac{\xi_{(\varphi)} \left(1-\xi^2 \right)}{\left[M+\sqrt{M^2-Q^2\left(1-\xi^2 \right)}\right] }}\,,
\label{33}
\end{equation}
which coincides with the meridional angle $\theta = \eta $; it can be used as a delimiter to distinguish some special value of the angle $ \theta  = \theta_{*}$.

\subsubsection{Third guiding function, $\zeta$}

Also, the spacetime with the metric (\ref{31}) admits two Killing vectors
$$
\xi^i_{(1)} {=} \sin{\varphi} \ \delta^i_{\theta} {+} {\rm ctg}{\theta} \cos{\varphi} \ \delta^i_{\varphi} \,,
$$
\begin{equation}
\xi^i_{(2)} {=} \cos{\varphi} \ \delta^i_{\theta} {-} {\rm ctg}{\theta} \sin{\varphi} \ \delta^i_{\varphi} \,.
\label{331}
\end{equation}
The difference of their squares
\begin{equation}
\xi^2_{(1)}-\xi^2_{(2)}  =  r^2 \sin^2{\theta} \cos{2\varphi}
\label{332}
\end{equation}
gives the hint for reconstruction of the third guiding function $\zeta$:
\begin{equation}
\zeta \equiv \frac12 {\rm arccos}\left[\frac{\xi^2_{(1)}-\xi^2_{(2)}}{\xi^2_{(\varphi)}} \right]\,,
\label{333}
\end{equation}
which coincides with the azimuthal angle $\varphi = \zeta$. This scalar can be used as a delimiter to distinguish some special value of the angle $ \varphi = \varphi_{*}$.

To conclude, one can say the following: first, when we intend to introduce on the invariant level the special sphere (e.g., to mark the dark matter wall) we have to use the first invariant $\xi$ and the delimiter $\xi=\xi_{*}$; second, when we intend to describe the straight dark matter filament, we can use two requirements $\eta {=} \theta_{*}$ and $\zeta {=} \varphi_{*}$ to fix the corresponding line. In this paper we will illustrate the idea to use  the scalar $\xi$ only; we plan to discuss the problem of filaments in the next work.

\subsection{Master equations of the axionic magneto-electro-statics}

\subsubsection{The key equation}

The equations (\ref{23}) for the electromagnetic field near the static spherically symmetric dyon are known to be reduced to one equation
\begin{equation}
\frac{d}{dr}\left(r^2 \frac{d A_0}{dr} + \mu \phi \right)  = 0  \,,
\label{34}
\end{equation}
where $A_0(r)$ is the electrostatic potential, and the magnetic charge $\mu$ is associated with the magnetic potential $A_{\varphi}= \mu (1{-}\cos{\theta})$ (see, e.g., \cite{BG1} for references).  Integration of (\ref{34}) yields
\begin{equation}
\frac{d A_0}{dr}  = \frac{{\cal Q}(r)}{r^2} \,, \quad  {\cal Q}(r) \equiv Q_* - \mu \phi(r)  \,,
\label{35}
\end{equation}
where ${\cal Q}(r)$ is the so-called effective charge, which depends on the axion field, and  $Q_*$ is the integration constant.
When we search for solutions continuous on the interval $r_{+}<r<\infty$, we can link the parameter ${\cal Q}(\infty) =Q_{*}{-}\mu \phi{(\infty)}$ with a total electric charge $q$ of the object, which could be found by a distant observer. This means that the quantity $-\mu \phi{(\infty)}$ plays the role of an effective axionically induced electric charge, and it is predetermined by the behavior of the pseudoscalar field at infinity. When we search for solutions continuous on the interval $r_{+}<r< r_{*}$, we have to link the constant $Q_{*}$ with the value of the pseudoscalar field $\phi(r_{*})$ on the delimiting surface $r=r_{*}$. Clearly, the electric potential can be found in quadratures, when the profile $\phi(r)$ is known:
\begin{equation}
 A_0(r)  = A_0(r_0) + Q_* \left(\frac{1}{r_0} - \frac{1}{r}\right) - \mu \int^r_{r_{0}} \frac{d\rho}{\rho^2} \ \phi(\rho)  \,.
\label{356}
\end{equation}
The equation for the axion field (\ref{24}) with the potential (\ref{21}) can be reduced to the form:
\begin{equation}
\frac{d}{dr} \left(r^2 N \frac{d \phi}{dr} \right) - \frac{m^2_{A} r^2 \Phi_*}{2\pi}  \sin{\left(\frac{2\pi \phi}{\Phi_*}\right)}= - \frac{\mu}{\Psi^2_0} \frac{d A_0}{dr} \,.
\label{36}
\end{equation}
Excluding $A_0$ from this equation using (\ref{35}), and applying the ansatz concerning the equilibrium state (i.e., $\phi {=} n \Phi_*$, $n \neq 0$),
we obtain the key equation of the model in terms of the variable $r$:
\begin{equation}
r^2 \frac{d}{dr}\left(r^2 N \frac{d \Phi_{*}}{dr}\right)  =  \frac{\mu^2}{\Psi^2_0}\left(\Phi_{*} - \frac{Q_*}{n \mu}\right) \,.
\label{38}
\end{equation}
Since $\xi = \sqrt{N(r)}$, we can rewrite the equation (\ref{38}) in terms of $\xi$ using the relationship
\begin{equation}
r = \frac{M + \sqrt{\xi^2 Q^2 + (M^2-Q^2)}}{1-\xi^2} \,.
\label{39}
\end{equation}
The sign in front of square root is chosen so that the outer horizon $r{=}r_{+}{=} M{+}\sqrt{M^2{-}Q^2}$ corresponds to the value $\xi{=}0$. We denote the derivative with respect to $\xi$ as a prime, and obtain the following differential equation:
\begin{equation}
\Phi_{*}^{\prime \prime}(\xi) {+}  \left(\frac{1}{\xi} {+} \frac{\xi}{\xi^2 {+} \nu} \right) \Phi_{*}^{\prime}(\xi) -  \frac{\mu^2 \left( \Phi_{*}{-} \frac{Q_*}{n\mu} \right)}{\Psi^2_0 Q^2 \left(\xi^2 {+} \nu \right)} =0 \,,
\label{40}
\end{equation}
with the guiding parameter $\nu$ given by
\begin{equation}
\nu = \frac{M^2{-}Q^2}{Q^2} \,.
\label{41}
\end{equation}
Below we indicate (\ref{40}) as the key equation.

\subsubsection{Stepwise equilibrium functions}

The ansatz about the equilibrium function $\phi = n \Phi_*(\xi)$ can be extended as follows. We assume now, that there are two domains $0<\xi<\xi_{*}$ and $\xi_{*}<\xi<1$, divided by the spherical surface indicated by the value $\xi_{*}$ of the scalar $\xi$. We assume that the integer $n$ takes the values $n_1$ and $n_2$ in the first and second domains, respectively.
The equilibrium function describing the pseudoscalar field can be now presented by the stepwise function
\begin{equation}
\phi = \phi_{(\rm eq)}= \Phi_*(\xi) \left[n_1 h(\xi_{*} -\xi) + n_2 h(\xi - \xi_{*}) \right] \,,
\label{42}
\end{equation}
where $h(z)$ is the Heaviside function, equal to one, when $z \geq 0$ and to zero, when $z<0$. Why this extension is considered to be interesting?
From the mathematical point of view, this extension keeps the fundamental properties of the potential (\ref{21}), i.e., the potential and its first derivative take zero values, when the axion field is in the equilibrium state. Indeed, we can easily check the following formulas:
$$
\frac{2\pi^2 V(\phi_{(\rm eq)})}{m^2_A \Phi^2_{*}} =  1{-} \cos{\left\{2\pi\left[n_1 h(\xi_{*} {-}\xi) {+} n_2 h(\xi {-} \xi_{*}) \right] \right\}}=
$$
$$
= 1{-}
\cos{\left[2\pi n_1 h(\xi_{*} -\xi)\right]} \cos{\left[2\pi n_2 h(\xi-\xi_{*})\right]} {+}
$$
\begin{equation}
{+} \sin{\left[2\pi n_1 h(\xi_{*} {-}\xi) \right]} \sin{\left[2\pi n_2 h(\xi{-}\xi_{*})\right]}
=0 \,,
\label{43}
\end{equation}
$$
\frac{\pi}{m^2_A \Phi_{*}}\frac{dV}{d\phi}(\phi_{(\rm eq)})  {=} \sin{\left\{2\pi\left[n_1 h(\xi_{*} {-}\xi) {+} n_2 h(\xi {-} \xi_{*}) \right]\right\}}{=}
$$
$$
=\sin{\left[2\pi n_1 h(\xi_{*} {-}\xi)\right]} \cos{\left[2\pi n_2 h(\xi-\xi_{*})\right]}
+
$$
\begin{equation}
+ \sin{\left[2\pi n_2 h(\xi-\xi_{*})\right]} \cos{\left[2\pi n_1 h(\xi_{*} {-}\xi) \right]} =
0 \,.
\label{44}
\end{equation}
In principle, the mentioned unique properties of the periodic potential allow us to extend the described procedure from the example of the two-level profile to a multi-level profiles.
In other words, one can consider three, four, etc. levels in the axionic dark matter profiles, using the extension of the formula (\ref{42}) for the set of integers $n_1$, $n_2$, $n_3$, $n_4$, etc.
As for the function $\Phi_{*}(\xi)$, its profile plays the role of a common envelope.

\subsubsection{Conditions on the boundary $\xi=\xi_{*}$}

For the two-level profiles one has to solve the key equation (\ref{40}) in two domains $0<\xi<\xi_{*}$ and $\xi_{*}<\xi<1$  separately, and to assume, that the parameter $Q_{*}$ takes  different values in these domains $Q_{*}^{(1)} \neq Q_{*}^{(2)}$. The envelope function $\Phi_{*}(\xi)$ is considered to be continuous near the surface $\xi=\xi_{*}$,
\begin{equation}
[\Phi_{*}] \equiv \lim_{\epsilon \to 0} \left\{\Phi_{*}(\xi_{*}+ \epsilon)- \Phi_{*}(\xi_{*}- \epsilon) \right\} = 0 \,.
\label{jump1}
\end{equation}
This is possible, when the parameters $Q_{*}^{(1)}$ and $Q_{*}^{(2)}$ are linked as follows:
\begin{equation}
\frac{Q_{*}^{(1)}}{n_1} = \frac{Q_{*}^{(2)}}{n_2} \equiv  \frac{Q_{*}}{n_*} \,,
\label{jump2}
\end{equation}
providing the key equation (\ref{40}) to be the same in both spatial domains.
When $n_1 \neq n_2$, we deal with the jump of the pseudoscalar field (\ref{42}) on the boundary $\xi=\xi_{*}$
\begin{equation}
[\phi] \equiv \lim_{\epsilon \to 0} \left\{\phi(\xi_{*}{+} \epsilon){-} \phi(\xi_{*}{-} \epsilon) \right\} = (n_2{-}n_1) \Phi_{*}(\xi_{*})  \,.
\label{jump3}
\end{equation}
The value of this jump $[\phi]$ is equal to zero, if the delimiting value $\xi_{*}$ coincides with the null of the envelope function, i.e., $\Phi_{*}(\xi_{*})=0$.
Similarly, the jump of the derivative $[\phi^{\prime}]$ vanishes, when two conditions are satisfied: $\Phi_{*}(\xi_{*})=0$ and $\Phi^{\prime}_{*}(\xi_{*})=0$.
Since the integer $n_1$ appears in the denominator (see (\ref{jump2})), the case $n_1=0$ should be considered separately; now the scheme of analysis is consistent, if we put $Q^{(1)}_{*}=0$.

From the physical point of view, the condition $[\phi] \neq 0$ means that we deal with a wall with a non-zero surface density of axions.
Modeling of the two-level distributions of the axionic dark matter is also interesting for description of profiles near black holes. We mean that a radius can exist, say $r_{*}$, which marks the specific boundary: $\xi=\xi_{*}$. When $\xi<\xi_{*}$ we find that $n_1=0$ and thus $\phi_{(\rm eq)} = 0$, i.e., all the dark matter particles are  absorbed by the black hole.
When $\xi>\xi_{*}$ the axionic dark matter profile is not empty, i.e. $n_2 \neq 0$, and particles rotating around the black hole can resist to the gravitational attraction.

\section{Analysis of the key equation}

\subsection{The Heun equation}

The key equation (\ref{40}) is the particular case of the known Heun equation
\begin{equation}
Y^{\prime \prime} + Y^{\prime} \left[ \frac{\epsilon}{x{-}a} {+} \frac{\delta}{x{-}1} {+} \frac{\gamma}{x} \right] +  Y \frac{\alpha \beta x {-} \rho}{(x{-}a)(x{-}1)x}  = 0   \,,
\label{Heun}
\end{equation}
(see, e.g., \cite{Heun1,Heun2}), which is in its turn the particular case of the Fuchs equation \cite{Ince,Poole}. The solution of this equation is regular at infinity, when $\epsilon{+}\gamma{+}\delta=\alpha{+}\beta {+}1$.
The equation (\ref{40}) can be transformed to (\ref{Heun}) using the relationship $x = \frac{i \xi}{\sqrt{\nu}}$, if we put
$$
\epsilon = \delta = \frac12 \,, \quad \gamma =1 \,, \quad a=-1 \,, \quad \rho =0 \,,
$$
\begin{equation}
 \alpha = \frac12 + \sqrt{\frac14 + \frac{\mu^2}{\Psi^2_0 Q^2}}  \,, \quad \beta = \frac12 - \sqrt{\frac14 + \frac{\mu^2}{\Psi^2_0 Q^2}}  \,.
\label{Heun34}
\end{equation}
The dimensionless guiding parameter $\nu {=} \frac{M^2{-}Q^2}{Q^2}$ plays essential role in the analysis of the key equation.

1) The standard model relates to the positive value $\nu>0$ (or  $M^2>Q^2$); in this case we deal with the object (e.g., the charged black hole), which has  the inner and outer horizons. The key equation (\ref{40}) is characterized by one real singular point only,  $\xi {=}0$, which is situated on the left boundary of the admissible interval for $\xi$.

2) When $\nu$ is negative, we deal with the so-called naked central singularity in the spacetime (singularity without horizons). In this case in the key equation there are two real singular points: $\xi=0$ and $\xi = \sqrt{|\nu|}$ (we consider $\xi$ to be positive).

3) The intermediate case $\nu{=}0$ (or $M^2{=}Q^2$) describes the so-called extremal black hole, in which outer and inner horizons coincide. Again, the key equation (\ref{40}) is characterized by one real singular point only,  $\xi {=}0$.
First of all, we consider in more detail this intermediate case.

\subsection{$M^2{=}Q^2$: Extremal black hole}

\subsubsection{General solution to the key equation}

When $M^2{=}Q^2$, the inner and outer horizons of the object coincide, and the equation (\ref{40}) can be reduced to the Euler equation
\begin{equation}
\xi^2 \Phi_{*}^{\prime \prime}(\xi) +  2\xi \Phi_{*}^{\prime}(\xi) -  \frac{\mu^2}{\Psi^2_0 M^2} \left(\Phi_{*}- \frac{Q_{*}}{n_{*}\mu} \right) =0   \,.
\label{46}
\end{equation}
The general solution to (\ref{46}) has the following form:
\begin{equation}
\Phi_{*}(\xi) = \frac{Q_{*}}{n_{*}\mu} + C_1 \xi^{\sigma_1} + C_2 \xi^{\sigma_2}   \,,
\label{47}
\end{equation}
where $C_1$  and $C_2$ are the constants of integration, and the power indices $\sigma_1$ and $\sigma_2$
$$
\sigma_1 = \frac12 \left(\sqrt{1+ \frac{4\mu^2}{\Psi^2_0 M^2}} -1 \right) \geq 0 \,,
$$
\begin{equation}
\sigma_2 = -\frac12 \left(\sqrt{1+ \frac{4\mu^2}{\Psi^2_0 M^2}} +1 \right) <0  \,,
\label{48}
\end{equation}
are of the opposite sign. Since $\sigma_1{+}\sigma_2={-}1$, we have only one parameter of modeling, which is associated, in fact, with the value of the ratio $\frac{\mu^2}{\Psi^2_0 M^2}$.

\subsubsection{The solution regular in the interval $\xi_0<\xi<1$}

Here we assume that at $\xi=\xi_{0}>0$  the function $\Phi_{*}(\xi)$ takes the value $\Phi_{*}(\xi_0)$. Such a boundary condition is typical for the case, when we deal with a magnetic star, and the parameter $\xi_{0}$ relates to its radius.  The spatial infinity $r \to \infty$ relates to $\xi \to 1$, and the corresponding value of the function $\Phi_{*}$ is indicated as  $\Phi_{*}(1) \equiv \Phi_{\infty}$. Then the solution for the function $\Phi_{*}(\xi)$ takes the analytical form
$$
\Phi_{*}(\xi) = \frac{Q_*}{n_{*}\mu}\left(1-\xi^{\sigma_2} \right) + \Phi_{\infty}\xi^{\sigma_2} +
$$
\begin{equation}
+ \left(\frac{\xi^{\sigma_1} {-} \xi^{\sigma_2}}{\xi^{\sigma_1}_0 {-} \xi^{\sigma_2}_0} \right)
\left[\Phi_*(\xi_0) {-} \frac{Q_*}{n_{*}\mu}\left(1{-}\xi^{\sigma_2}_0 \right) {-} \Phi_{\infty} \xi_0^{\sigma_2}\right]
\,.
\label{49}
\end{equation}
There are two guiding parameters in this formula: the amplitude factor $\frac{Q_*}{n_{*}\mu}$, which can be positive or negative with respect to the signs  of the constants $Q_*$ and $\mu$, as well as, the positive factor $\frac{\mu^2}{\Psi^2_0 M^2}$. Typical behavior of these profiles is illustrated in Fig.1.

\begin{figure}
	\includegraphics[width=90mm,height=70mm]{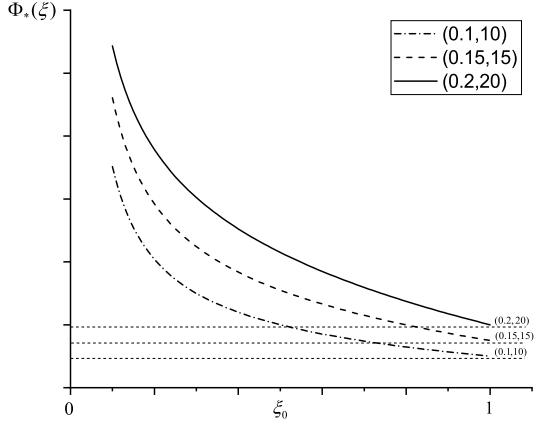}
	\caption {Typical profiles of the regular envelope function $\Phi_*(\xi)$ for the case of extremal charged black hole ($Q^2{=}M^2$) given by the formula (\ref{49}).  The first number in the parentheses relates to the value of the parameter $\frac{\mu^{2}}{\Psi_{0}^{2}M^{2}}$, the second one corresponds to the parameter $\frac{Q_{*}}{n\mu}$.}
\end{figure}

\subsubsection{An example of the stepwise solution}

Let us consider the illustration, which corresponds to the equilibrium function
(\ref{42}) with the delimiter $\xi_{*}{=}\frac14$, and the model parameters $n_1{=}0$, $n_2{=}1$, $Q^{(1)}_{*}{=}0$
\begin{equation}
\phi_{(\rm eq)}= \Phi_{*}(\xi) \times h \left(\xi - \frac14\right) \,,
\label{5231}
\end{equation}
where the envelope function in the interval $\frac14 <\xi<1$ is of the form
\begin{equation}
\Phi_{*}(\xi) = \frac{1}{81}\left(\frac{1}{\xi^2} {+} 128 \xi {-} 48 \right) \,.
\label{5235}
\end{equation}
This envelope function takes zero value at $\xi=\frac14$; its derivative
\begin{equation}
\Phi^{\prime}_{*}(\xi) =  \frac{2}{81}\left(64{-}\frac{1}{\xi^3} \right)
\label{5132}
\end{equation}
also vanishes at  $\xi=\frac14$. This function is the exact solution to the key equation with the following values of the model parameters:
\begin{equation}
\frac{Q^{(2)}_*}{n_2 \mu}= {-} \frac{16}{27} \,,  \quad \frac{\mu}{\Psi_0 M} =\sqrt2 \,, \quad  \sigma_1=1 \,,  \sigma_2=-2 \,.
\label{5033}
\end{equation}
and takes the value $\Phi_{\infty}=1$ at the spatial infinity. Mention should be made that $\Phi(\xi)=0$ is the exact solution to the key equation in the interval $0<\xi<\frac14$. On the delimiting sphere $\xi=\frac14$ the function (\ref{5235}) and the constant function $\Phi(\xi)=0$ happen to be sewed.
The profile of the function (\ref{5231}) with (\ref{5235}), as well as, two additional ones are depicted in Fig.2.

\begin{figure}
	\includegraphics[width=90mm,height=70mm]{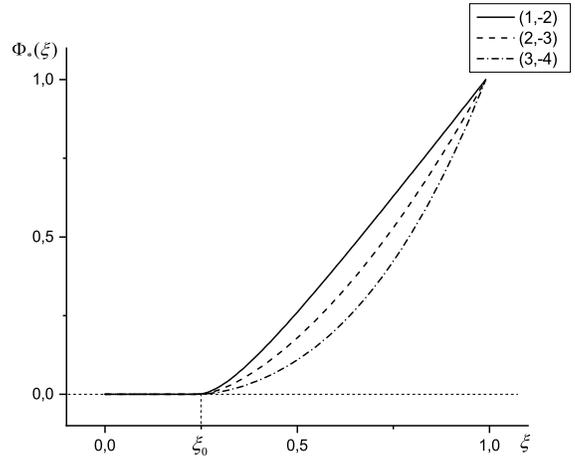}
	\caption {Three examples of the stepwise envelope function (\ref{5231}); the profiles are distinguished by the parameters $\sigma_{1}$ and $\sigma_{2}$, indicated in the parentheses.
 Graphs of all the functions have the delimiter $\xi_0 = \xi_* = \frac14$, and tend to one at infinity.}
\end{figure}

\subsection{$Q=0$: The Schwarzshild type black hole}

\subsubsection{Key equation and its general solution}

When $Q^2<<M^2$, the geometry of the background spacetime is close to the one of the Schwarzshild type; we mean that the magnetic charge $\mu$ is non-vanishing and it plays an important role in the axionic halo formation, however, in the formation of the background gravitational field its contribution is negligible. Now we can obtain the reduced key equation from (\ref{40}) as the limiting case $Q \to 0$; we deal now with the Bessel equation
\begin{equation}
\xi^2 \Phi_{*}^{\prime \prime}(\xi) +  \xi \Phi_{*}^{\prime}(\xi) -  \frac{\mu^2}{\Psi^2_0 M^2} \ \xi^2 \left(\Phi_{*} - \frac{Q_{*}}{n_{*}\mu} \right) =0   \,.
\label{53}
\end{equation}
The general solution to (\ref{53}) is
\begin{equation}
\Phi_{*}(\xi) = \frac{Q_{*}}{n_{*}\mu} + C_1 I_0\left(\frac{\mu \xi}{\Psi_0 M}\right) + C_2 K_0\left(\frac{\mu \xi}{\Psi_0 M} \right)   \,.
\label{54}
\end{equation}
Here $I_0$ and $K_0$ are the Bessel functions of the third and fourth type with indices equal to zero; they can be represented standardly as follows:
\begin{equation}
I_0(z) = \sum_{m=0}^{\infty}\frac{\left(\frac{z}{2}\right)^{2m}}{(m!)^2} \,, \quad I_0(z \to 0) \approx 1 + \frac{z^2}{4} \,,
\label{55}
\end{equation}
\begin{equation}
K_0(z) = \int_0^{\infty} d \zeta e^{-z \cosh{\zeta}} \,, \quad K_0(z \to 0) \approx \log{\frac{z}{2}} \,.
\label{56}
\end{equation}

\subsubsection{Regular solution in the interval $0<\xi<1$}

If we consider the requirement $\Phi_{*}(0) = 0$ we find immediately the regular profile
\begin{equation}
\Phi_{*}(\xi) = \frac{Q_{*}}{n_{*} \mu} \left[1 - I_0\left(\frac{\mu \xi}{\Psi_0 M} \right) \right] \,.
\label{57}
\end{equation}
Clearly, $\Phi^{\prime}_{*}(0)=0$ and $\Phi_{*}(1) = \frac{Q_{*}}{n_{*} \mu} \left[1 - I_0\left(\frac{\mu}{\Psi_0 M} \right) \right]$.
In other words, the values of the function $\Phi_{*}(\xi)$ and of its derivative are equal to zero at the Schwarzschild horizon, and the maximal value of the modulus of this function depends on the following two ratios:  $\left|\frac{Q_{*}}{n_{*} \mu}\right|$ and $\Gamma \equiv \frac{\mu}{\Psi_0 M}$.

\subsubsection{Stepwise equilibrium function}

Again we assume that $n_1=0$, $n_2=1$ and $Q^{(1)}_{*}=0$, so that the stepwise equilibrium function is of the form
\begin{equation}
\phi_{(\rm eq)}= \Phi_{*}(\xi) \times h \left(\xi - \xi_{*}\right) \,,
\label{7231}
\end{equation}
with the envelope function
\begin{equation}
\Phi_*(\xi) {=} \frac{Q_*}{n_* \mu} \left\{1 {-} \frac{K^{\prime}_0(\Gamma \xi_*) I_{0}(\Gamma \xi) {-} K_0(\Gamma \xi) I^{\prime}_{0}(\Gamma \xi_*)}{K^{\prime}_0(\Gamma \xi_*) I_{0}(\Gamma \xi_*) {-} K_0 (\Gamma \xi_*) I^{\prime}_0(\Gamma \xi)} \right\},
\label{599}
\end{equation}
which is the exact solution to the key equation. The prime denotes the derivative with respect to the argument of the function. We introduced here the auxiliary parameter $\Gamma {=} \frac{\mu}{\Psi_0 M}$. Clearly, $\Phi_*(\xi_*) {=}0$ and $\Phi^{\prime}_*(\xi_*) {=}0$. The delimiter value $\xi_*$ satisfies the transcendent equation $\Phi_*(1)= \Phi_{\infty}$. The typical behavior of the envelope function is presented in Fig.3.

\begin{figure}
	\includegraphics[width=90mm,height=70mm]{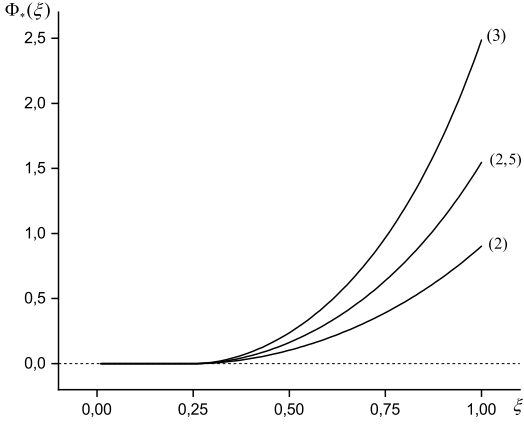}
	\caption {Illustrations of the stepwise equilibrium function (\ref{7231}) with the envelope function (\ref{599}). For all profiles $\frac{Q_{*}}{n_{*}\mu}{=}1$ and the delimiter is $\xi_*{=}0.25$. The values of the parameter $\Gamma {=} \frac{\mu}{\Psi_0 M}$ are written in parentheses.}
\end{figure}

\subsection{$M=0$: The spacetime with naked singularity}

When $M{=}0$ the spacetime has no horizons, and the variable $\xi$ takes values in the interval $(1,\infty)$.
Now $\nu{=}{-}1$, thus the equation (\ref{40}) can be reduced to the Heun equation (\ref{Heun}) with the parameters (\ref{Heun34}) by the replacement $\xi \to x$. Now the key equation has no singular points in the interval $(1,\infty)$. Illustrations are presented in Fig.4.
\begin{figure}
	\includegraphics[width=90mm,height=70mm]{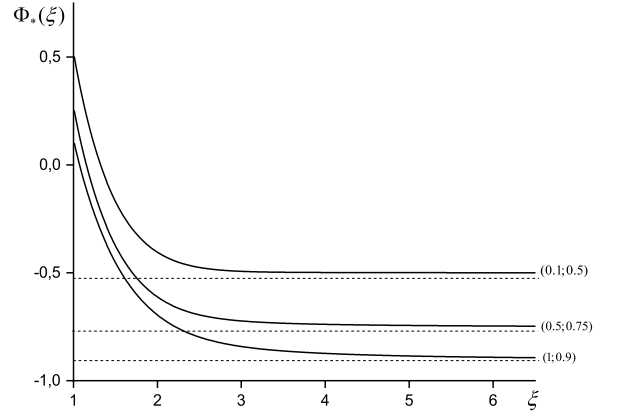}
	\caption {Typical profiles of the envelope function $\Phi_*(\xi)$ for the magnetic naked singularity as the solutions to the Heun equation (\ref{40}). The scalar $\xi$ takes values in the interval $(1,\infty)$; the point $\xi{=}1$ corresponds to the spatial infinity, and $\xi \to \infty$, when $r \to 0$. In the parentheses two parameters are presented: the first is  $\frac{\mu^{2}}{\Psi_{0}^{2}Q^{2}}$, the second is $\frac{Q_{*}}{n\mu}$. Asymptotically, at $\xi \to \infty$, the solutions to the Heun equation tend to the constants $\Phi_{0}$, which depend on the values of the chosen model parameters.}
\end{figure}

\subsection{Numerical analysis of the general case $M^2>Q^2$}

For illustration of the results of the numerical analysis of the case $\nu >0$, the most physically motivated, we studied systematically the Heun equation (\ref{40}) for various values of the guiding parameters $\frac{M^2{-}Q^2}{Q^2}>0$ and $\frac{\mu^2}{\Psi^2_0 Q^2}$. Now, we deal with the interval  $0<\xi<1$, and for the outer zone of the object there are no real singular points in the key equation (\ref{40}). Typical profiles of the envelope function $\Phi_*(\xi)$ are presented in Fig.5.
\begin{figure}
	\includegraphics[width=90mm,height=70mm]{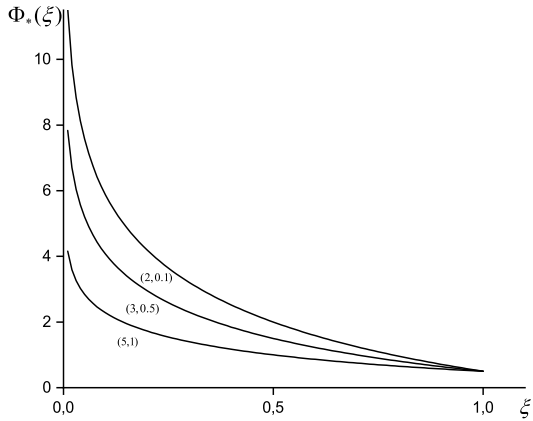}
	\caption {Typical profiles of the envelope function $\Phi_*(\xi)$, the solutions to the key equation (\ref{40}), for the case $M^2>Q^2$ and $\Phi(1){=}0.5$; the numbers in parentheses correspond to the values of the parameters $\frac{M^2}{Q^2}{-}1$ and $\frac{\mu^{2}}{\Psi_{0}^{2}M^{2}}$, respectively.}
\end{figure}

\section{Discussion and conclusions}

The authors of the work \cite{BEC} have formulated the idea that the dark matter axions form a Bose-Einstein condensate, and thus the behavior of the axion systems differs from the one of an ordinary cold dark matter, especially if the regime of interaction is non-linear and the external fields are strong. This idea emphasizes that the axion system is not a simple collisionless, pressureless cold gas; the axion system has to be characterized by some internal self-interaction.
We also follow the idea that axionic systems are self-interacting, and we think that their internal structures are predetermined by the modified periodic potential (\ref{21}). Minima of this potential predetermine the equilibrium states of the axions. Our modified periodic potential can be obtained from the standard one by introduction of the so-called envelope function $\Phi_{*}$, which stands to emphasizes the fact that the equilibrium value of the axion field depends on the position in the spacetime.

Since the strict variation formalism requires the envelope function to depend on some scalar invariants, $\Phi = \Phi_{*}(\xi,\eta,\zeta)$, we have linked these scalars with  moduli of three Killing vectors, which are associated with static spherically symmetric spacetime under consideration.

This extension of the axionic potential, in its turn, led us to the necessity of the Lagrangian modification, in which we considered new terms associated with the Killing vector field. Based on analogy with the Einstein-aether theory, we obtained the correspondingly extended master equations, thus providing the whole model to be self-consistent. The main conclusion of this first part of the work is the following: when the axion field is in the equilibrium state, for which the modified potential and its first derivative vanish, the presence of dynamically defined Killing vector fields does not violate the master equations for gravitational, electromagnetic and axion fields.

In Section III we considered the application of the developed formalism for the spacetime of the Reissner-Nordstr\"om type and have found exact solutions, which describe axionic halo profiles near the dyons and magnetic black holes. We have found the envelope functions for several values of guiding model parameters; the most interesting findings, from our point of view, are the solutions of the stepwise type, which describe two-level distributions of the axionic dark matter. In principle, such model distributions can be considered in the context of description of the magnetic black holes. Indeed, because of the gravitational attraction the axionic  dark matter halo surrounding such a object should have empty zone from the outer horizon till to the first stable orbit of the rotating massive particle (see Figs.2,3).

The developed formalism also can be applied to the description of the dark matter filaments; for this purpose we can use the scalars $\eta$ (\ref{33}) and $\zeta$ (\ref{333}). We hope to consider these cosmological units in the nearest future.

\begin{acknowledgments}
The work was supported by Russian Science Foundation (Project No. 16-12-10401), and, partially, by the Program of Competitive Growth of Kazan Federal University.
\end{acknowledgments}

\end{document}